# A Geometrically Frustrated Trimer-Based Mott Insulator

Loi T. Nguyen[1], T. Halloran[2], Weiwei Xie[3], Tai Kong[1], C.L. Broholm[2] and R.J. Cava[1]

[1]Department of Chemistry, Princeton University, Princeton, New Jersey 08544, USA

[2]Institute for Quantum Matter and Department of Physics and Astronomy,

Johns Hopkins University, Baltimore MD 21218, USA

[3]Department of Chemistry, Louisiana State University, Baton Rouge, Louisiana 70803, USA

**Abstract**

The crystal structure of $Ba_4NbRu_3O_{12}$ is based on triangular planes of elongated $Ru_3O_{12}$ trimers oriented perpendicular to the plane. We report that it is semiconducting, that its Weiss temperature and effective magnetic moment are -155 K and 2.59 $\mu_B$/f.u. respectively, and that magnetic susceptibility and specific heat data indicate that it exhibits magnetic ordering near 4 K. The presence of a high density of low energy states is evidenced by a substantial Sommerfeld-like T-linear term ($\gamma = 31(2)$ mJ/mole-$K^2$) in the specific heat. Electronic structure calculations reveal that the electronic states at the Fermi Energy reside on the $Ru_3O_{12}$ trimers and that the calculated density of electronic states is high and continuous around the Fermi Energy - in other words density functional theory calculates the material to be a metal. Our results imply that $Ba_4NbRu_3O_{12}$ is a geometrically frustrated trimer-based Mott insulator.

## 1. Introduction

The magnetic properties of insulating solids are typically governed by interactions between unpaired electronic spins located in the *d* or *f* orbitals of the constituent atoms.[1],[2] These interactions most frequently lead to ordered magnetic states at low temperatures. In some magnetic materials, however, all pair-wise magnetic spin interactions cannot be simultaneously satisfied due to the lattice geometry, frustrating the long-range magnetic order so that it occurs at temperatures significantly below that of the Weiss temperature $\Theta_W$, a measure of the average pair-wise magnetic interaction strength. This phenomenon is called geometric magnetic frustration and is most frequently observed in materials where the magnetic geometry is based on triangles. An experimental benchmark often used to indicate its presence in a material, whose universal application is under discussion[3],[4], is when the ratio of the Weiss temperature to the magnetic ordering temperature exceeds approximately $|\Theta_W|/T_M = 10$[5]. Materials that DFT calculations predict to be metallic conductors with significant density of electronic states at the Fermi Energy but are in fact electrically insulating are called Mott Insulators.[6],[7],[8] A non-interacting electron picture does not successfully describe these systems.

Ruthenium oxides display many unusual magnetic and electrical properties due to the multiple degrees of freedom of the Ru 4*d* electrons. $Sr_2RuO_4$, for example, becomes a superconductor below 1 K while $SrRuO_3$ is ferromagnetic at 160 K[9],[10] $CaRuO_3$ and $SrRuO_3$ have been characterized as Hund's rule metals.[11] Similar to the material described here, $Ba_4LnRu_3O_{12}$ (Ln = Lanthanide) hexagonal perovskites have been previously reported.[12],[13] In $Ba_4LaRu_3O_{12}$, for example, where La is nonmagnetic and $Ru_3O_{12}$ trimers are found, an antiferromagnetic transition is observed at 6 K. In related dimer-based materials[14], frustrated spin ½ magnetism has been reported for $Ba_3YRu_2O_9$.[15] The search for an alternative magnetic state based on the hexagonal perovskite geometry is the motivation for the current work.

Ba$_4$NbRu$_3$O$_{12}$ is reported to be isostructural with Ba$_4$HoRu$_3$O$_{12}$ and related compounds. Its atomic position coordinates have not been reported, and its experimentally determined magnetic properties have only been reported in a perfunctory fashion.[16] Here we show that Ba$_4$NbRu$_3$O$_{12}$, where the magnetic geometry is based on triangles, displays the classic signatures of frustrated magnetism, with the difference between it and the usual insulating magnetic materials being that the magnetic moment originates in Ru$_3$O$_{12}$ trimers rather than in individual RuO$_6$ octahedra (Ba$^{2+}$ and Nb$^{5+}$ are not magnetic)[17]. Ba$_4$NbRu$_3$O$_{12}$, with no apparent structural disorder and a Weiss temperature of − 155 K, shows evidence for some type of magnetic ordering upon cooling below about 4 K. The corresponding frustration index │Θ$_W$│ /T$_M$ of more than 30 is much larger than the benchmark value of 10. Resistivity measurements show that the material is semiconducting with a transport gap of 0.22 eV. First-principles calculations of the electronic structure further show that the electrons at the Fermi Energy are localized on the Ru$_3$O$_{12}$ trimers, shared by both Ru and O, and that in the one-electron standard DFT picture the material should be an electronic conductor. Thus, we conclude that trimer-based Ba$_4$NbRu$_3$O$_{12}$ embodies both Mott insulating behavior and geometric magnetic frustration in a single material.

## 2. Methods

Polycrystalline samples of Ba$_4$NbRu$_3$O$_{12}$ were synthesized by solid-state reaction using BaCO$_3$, RuO$_2$, Nb$_2$O$_5$ and Ru (Alfa Aesar, 99.99). First the precursor Ba$_5$Ru$_2$O$_{10}$ was prepared from BaCO$_3$ and RuO$_2$ on heating in air at 1000 °C for 12 hours. Next, the Ba$_5$Ru$_2$O$_{10}$ precursor, RuO$_2$, Nb$_2$O$_5$ and Ru powder were mixed thoroughly in the appropriate ratio, placed in an alumina crucible, and heated in air at 1000 °C for 12 hours. The resulting powder was re-ground, pressed into a pellet and heated in air at 1100 °C for 12 hours, and then at 1300 °C for 12 hours. The phase purity and crystal structure were determined through powder X-ray diffraction (PXRD) using a Bruker D8

Advance Eco with Cu Kα radiation and a LynxEye-XE detector. The structural refinements were performed with *GSAS*.[18] Crystal structure drawings were created by using the program *VESTA*.[19]

The magnetic susceptibility of $Ba_4NbRu_3O_{12}$ was measured by a Quantum Design Physical Property Measurement System (PPMS) DynaCool equipped with a VSM option. For those measurements, the samples were ground into powder and placed in plastic capsules. The magnetic susceptibility of $Ba_4NbRu_3O_{12}$, defined as *M/H*, where *M* is the sample magnetization and *H* is the applied field, was measured at the field of $H = 1$ kOe from 1.8 K to 300 K. The resistivity of $Ba_4NbRu_3O_{12}$ was measured by the DC four-contact method in the temperature range 250 K to 330 K with the PPMS. The sample was pressed, sintered, and cut into pieces with the approximate size $1.0 \times 2.5 \times 1.5$ mm$^3$. Four Pt contact wires were connected to the sample using silver paint. The specific heat was measured by the adiabatic heat pulse method in a Dynacool PPMS equipped with a dilution refrigerator option from 0.1 K to 200 K.

The single electron band structure of $Ba_4NbRu_3O_{12}$ was calculated using CAESAR[20] with semi-empirical extended-Hückel-tight-binding (EHTB) methods[21] and the following modified parameters for Ba: $\zeta = 5.486$, $H_{ii} = -7.00$ eV; 6*p*: $\zeta = 3.559$, $H_{ii} = -4.80$ eV. To verify these results, the electronic structure of $Ba_4NbRu_3O_{12}$ including the band structure and electronic Density of States (DOS) was also calculated using WIEN2k in the local density approximation (LDA). Spin-orbit coupling (SOC) was included in the calculation. The k-point mesh was set at 10*10*10. Spin polarization using LDA+S was employed in the ferromagnetic and antiferromagnetic models investigated.

## 3. Results

### 3.1 Crystal Structure

The powder X-ray diffraction pattern and structural refinement of $Ba_4NbRu_3O_{12}$ are shown in Figure 1. The structure of $Ba_4HoRu_3O_{12}$ was used as the starting model.[13] $Ba_4NbRu_3O_{12}$ crystallizes in a rhombohedral structure with the space group *R-3m* (No.166). The lattice parameters of $Ba_4NbRu_3O_{12}$ are in agreement with the previous report[16] and the structural parameters are summarized in Table 1. As seen in Figure 2, the structure of $Ba_4NbRu_3O_{12}$ is based on three $RuO_6$ octahedra connected by face-sharing to form $Ru_3O_{12}$ trimers. The average distance between Ru atoms in the $Ru_3O_{12}$ trimers is quite short, 2.51 Å, due to face sharing, an indication of strong chemical interactions between them. The Ru-O distances are 2.016(6) Å and 2.045(7) Å in the outer $RuO_6$ octahedra and 2.028(6) Å in the inner $RuO_6$ octahedron, suggesting that the Ru's are electronically equivalent.[22],[23] The $NbO_6$ octahedra and the $Ru_3O_{12}$ trimers alternate, connected by corner-sharing, to generate the "12 layer" (12L, three times 4 layers of $MO_6$ octahedra) hexagonal perovskite type structure of this material. The layers are staggered such that if each trimer is considered as a single magnetic entity, then an individual trimer has 3 trimers in the layer above and 3 in the layer below in the geometry of an octahedron. A schematic view of the arrangement of the $Ru_3O_{12}$ trimers in $Ba_4NbRu_3O_{12}$ is drawn in Figure 3a. The trimers are centered at z = ⅙, ½ and ⅚. No structural disorder was detected by our methods.

**3.2 Magnetic Properties**

The temperature-dependent magnetic susceptibility of $Ba_4NbRu_3O_{12}$ and its reciprocal are plotted in Figure 4. The magnetic data for $Ba_4NbRu_3O_{12}$ from 100-300 K are well fit by the Curie-Weiss law $\chi = \frac{C}{T-\theta_W} + \chi_o$, where $\chi_o$ is the temperature-independent part of the susceptibility, *C* is the Curie constant and $\Theta_W$ is the Weiss temperature. Through least squares fitting, these are found to be -155 K, 2.59 $\mu_B$/f.u., and 5.25×10⁻⁴ emu Oe⁻¹ mol-f.u.⁻¹ respectively. The temperature-dependent magnetic susceptibility of $Ba_4NbRu_3O_{12}$ under the field 1 kOe shows only a little downward curvature at the lowest temperatures measured.

Because both $Ba^{2+}$ and $Nb^{5+}$ are nonmagnetic, the intertrimer and intratrimer interactions of the $Ru_3O_{12}$ units determine the magnetic properties of $Ba_4NbRu_3O_{12}$. The number of $4d$ electrons in the trimer is 13, an odd number, (The average oxidation state of Ru is +3.67) and thus in a Ru-only (i.e. $Ru_3$) model the electronic configuration of the $Ru_3O_{12}$ trimer[24], with $D_{3d}$ point symmetry, is expected to be $(a_{1g})^2(e_g)^4(a_{2u})^2(e_u)^4(e_g)^1$ (Figure 3b). In this picture there is one unpaired electron per trimer, and hence the expected spin of the trimer would be S = ½; the resulting effective moment would be a $\mu_{eff}$ of $g\sqrt{\{s(s+1)\}} \approx 1.73$ $\mu_B$ per formula unit. A spin of 3/2 per trimer (approximately 3.87 $\mu_B$ per formula unit) is also possible in a localized spin picture, as is S = 5/2 or 5.92 $\mu_B$/f.u (depending on how the spins in the individual octahedra are oriented with respect to each other). Our experimentally observed effective moment (2.59 $\mu_B$/formula unit) is larger than the S = 1/2 value, and smaller than the S = 3/2 and S = 5/2 values. We note that in the related material $Ba_4LaRu_3O_{12}$, with 11 electrons per trimer, a moment of 2.85 $\mu_B$/f.u is observed, which, as in the current case, is different from what is expected for a simple S = ½ dimer ground state[15],[25] although the temperature dependence of the susceptibility and the signature of magnetic ordering are different from what is observed here. Finally, the inset in Figure 4 shows the magnetization of $Ba_4NbRu_3O_{12}$ as a function of applied field up to 9 T at 1.8 K and 300 K. At 300 K, there is a linear relation between the magnetization and the applied field. This trend becomes curved at 1.8 K, but with no sign of saturation up to the highest field applied. The onset field for the curvature in M vs. H is well beyond the 1000 Oe field employed to measure the temperature-dependent susceptibility.

Figure 5 shows the field cooled (FC) and zero field cooled (ZFC) DC susceptibility in an applied field of 100 Oe for $Ba_4NbRu_3O_{12}$. The bifurcation between FC and ZFC susceptibility near 4 K indicates the onset of some type of frozen spin state, although its microscopic character is not currently known. Because we do not see any disorder between Nb and Ru in the structural refinement, this suggests that geometric magnetic frustration in $Ba_4NbRu_3O_{12}$ is key to this state even if undetected disorder also plays a role. Similar bifurcation between FC and ZFC susceptibility was seen in $Ba_4TbRu_3O_{12}$ at 24 K (Tb is a magnetic rare earth) where an anomaly in the specific heat at 24 K was taken to indicate the presence of long range antiferromagnetic ordering.[13]

Figures 6 and 7 show the specific heat characterization of $Ba_4NbRu_3O_{12}$ over a wide temperature range. No non-magnetic analog of this material is available, and so we used least squares fitting to the high $T$ phonon-dominated specific heat to estimate and then subtract $C_{phonon}(T)$ from the total specific heat and isolate $C_{mag}(T) = C(T) - C_{phonon}(T)$. We constrained the fit as follows: (1) We require $C_{phonon}(T) \leq C(T)$ throughout. (2) In the absence of dimerization and spin order, the smallest possible spin-orbital degeneracy that can be associated with the $Ru_3O_{12}$ unit is 2 so we constrain the total magnetic entropy to be $\Delta S_{mag}(T) = \int (C_{mag}/T)dT = R\ln 2$. The effective moment inferred from susceptibility data exceeds that of a spin-1/2 degree of freedom so the magnetic entropy may exceed this value. The lower $T$ features that we shall focus on are however, not strongly dependent on this constraint. (3) We use a functional form for the non-magnetic specific heat that is consistent with a harmonic phonon system: A superposition of two Debye spectra and an Einstein mode is needed to achieve a fit to the phonon specific heat in the temperature range from 15 K to 200 K that is consistent with constraints (1-2). The corresponding phonon model (Figure 6a and Table II) converges towards a high $T$ specific heat of 58(3)$R$, which is consistent with the Dulong-Petit value of $3nR$, where $n=20$ is the number of atoms per formula unit. Figure 6b shows convergence of the magnetic entropy to $R \ln 2$ imposed by the constraint. While a definite determination of $C_{mag}(T)$ must await a separate measurement of the phonon density of states or the discovery of a non-magnetic reference sample, our present determination of $C_{mag}(T)$ (Figure 6c-d and Figure 7) is independent of the phonon model at lower $T$. The higher $T$ peak in $C_{mag}(T)$ (Figure 6c) indicates buildup of short range magnetic correlations on the temperature scale set by $\Theta_{CW}$ while the lower $T$ peak is associated with spin freezing and the bifurcation point in the FC/ZFC susceptibility. The limiting low $T$ form of the specific heat is independent of the detailed phonon model and is examined in Figure 6d and Figure 7. At the very lowest temperatures there is a small upturn in $C_{mag}(T)/T$ that we associate with the spin-full nuclear isotopes $^{99}$Ru and $^{101}$Ru with a combined natural abundance of 30%. Hyperfine coupling to the frozen ruthenium-based nuclear magnetism gives rise to a term of the form $C_{mag}(T) = B/T^2$. Beyond this term, Figure 6d indicates a finite value of $C_{mag}(T)/T$ as $T \rightarrow 0$. Figure 7 shows $C_{mag}(T)$ in a double logarithmic plot. Here weak curvature is visible that indicates a cross over $T \approx 0.4$ K between different power-law regimes. Overall, $C_{mag}(T) = \frac{B}{T^2} + \gamma T + AT^\alpha$ accounts well for all data for $T < 3$ K as shown with a solid line and decomposed as dashed lines in Figure 6d and Figure 7. The $T$-linear term in the specific heat $\gamma = 31(2)$ mJ/mole-K$^2$ has been seen in some geometrically frustrated spin-glasses and may arise from a random singlet state[26] or a spinon-fermi surface[27]. The additional power law term with $\alpha = 1.62(1)$ contrasts with observations in other quasi-two-dimensional geometrically frustrated magnets where $C_{mag}(T) \propto T^2$ has been associated with relativistic quasi-particles in two dimensions[28]. While it may not be possible to make a specific association with a collective quasi-particle here, our data and analysis establish the limiting behavior of the low energy magnetic density as a challenge for theory.

Finally, the resistivity of $Ba_4NbRu_3O_{12}$ is plotted as a function of reciprocal temperature in Figure 8. The data from 250-300 K was fit to the standard model: $\rho = \rho_o e^{\frac{E_a}{k_b T}}$ with an activation energy $E_a = 0.22$ eV. The inset shows the increase of resistivity upon cooling. With a resistivity activation energy of 0.22 eV, $Ba_4NbRu_3O_{12}$ is a semiconductor, as are other trimer-based compounds with different electron counts in the $Ba_4LnRu_3O_{12}$ (Ln=Lanthanide) family.[13], [15]

### 3.3 Electronic structure

Extended Hückel theory using Slater-type zeta functions was employed for the bonding analysis. Figure 9 illustrates in real space the highest occupied molecular orbitals (HOMOs) and lowest unoccupied molecular orbitals (LUMOs) for $Ba_4NbRu_3O_{12}$, which provide detailed information about the wave function of electrons near the Fermi level. The figure shows that they are localized on the $Ru_3O_{12}$ trimer. Further, there is no electron density between trimers (i.e. there is no orbital overlap), which, combined with the experimental result, indicates minimal electron hopping between trimers. The orbitals on the Ru and O atoms are strongly hybridized, indicating that the $Ru_3O_{12}$ trimer acts as the electronic and magnetic building block in this material. Finally, Ru2-Ru1-Ru2 bonding interactions exist within the trimers.

To further confirm this bonding analysis, the electronic band structure of $Ba_4NbRu_3O_{12}$ shown in Figure 10 was calculated using WIEN2k. Four different models including non-magnetic (NM), ferromagnetic (FM), antiferromagnetic (AFM-I) with $Ru_2\uparrow$-$Ru_1\downarrow$-$Ru_2\uparrow$, and antiferromagnetic-II (AFM-II) with $Ru_2\uparrow$-$Ru_1\downarrow$-$Ru_2\uparrow$, were examined, including their total energy, band structure and density of states. None of the electronic structures for the four models shows any band gap around the Fermi level. A half-filled narrow band straddles $E_F$ in the calculation. Thus, given the resistivity behavior of the material, the calculation results suggest that $Ba_4NbRu_3O_{12}$ is a Mott insulator.

## 4. Conclusions

Ba$_4$NbRu$_3$O$_{12}$ crystallizes in the 12-layer hexagonal perovskite unit cell with the *R*-3*m* space group. The material has high, activated resistivity, indicating that it is semiconducting. The large negative Weiss temperature and magnetic ordering near 4 K indicate that it is a geometrically frustrated magnet. The specific heat data indicate that a high density of low energy states is retained in the low T limit that could arise from the spinon fermi surface of a quantum spin liquid, from a random singlet state, or from semi-classical soft modes of a spin-glass. Further, the degeneracy of the highest filled orbital in the molecular picture for the Ru$_3$O$_{12}$ trimers suggests that orbital ordering is possible in this material. To distinguish these, and investigate the possible orbital ordering, a range of structural and spectroscopic work will be needed including inelastic magnetic neutron scattering. The calculation of a magnetic Hamiltonian would also be of interest. The semiconducting electrical properties contrast with the calculated metallic electronic structure. Our characterization suggests that Ba$_4$NbRu$_3$O$_{12}$ is a trimer-based Mott insulator that may present new opportunities for electronic doping of a geometrically frustrated magnet.

## 5. Acknowledgements

The authors are grateful to Zhijun Wang for his insights into the electronic structure of this material. This work was supported by the Basic Energy Sciences of the Department of Energy grant number DE-FG02-08ER46544 through the Institute for Quantum Matter at Johns Hopkins University, and at Louisiana State University startup funding.

**Figures**

Figure 1: Rietveld powder X-ray diffraction refinement of the structure of $Ba_4NbRu_3O_{12}$ in space group $R\text{-}3m$. The observed X-ray pattern is shown in black, calculated in red, difference ($I_{obs}-I_{calc}$) in blue, the background in green, and tick marks denote allowed peak positions in pink. $R_p$ = 0.0833, $R_{wp}$ = 0.1197, $\chi^2$ = 1.96.

Figure 2: The crystal structure of $Ba_4NbRu_3O_{12}$. The $NbO_6$ octahedra (dark green) are corner-shared with the $Ru_3O_{12}$ (purple) trimers made from face-shared $RuO_6$ octahedra, to form what is frequently referred to as a "12-layer" hexagonal perovskite structure. Barium is green, and oxygen is red.

Figure 3: (a) Schematic side view of the triangular layers of $Ru_3O_{12}$-trimers (the blue lozenges) in $Ba_4NbRu_3O_{12}$. The triangular layers of trimers are centered at z = ⅙, ½ and ⅚ in the unit cell. (b) The schematic energy of the molecular orbitals in the $Ru_3O_{12}$ trimers if only the Ru-Ru-Ru states are considered.

Figure 4: Main Panel - The temperature dependence of the magnetic susceptibility and the inverse of the difference between the magnetic susceptibility and the temperature independent magnetic susceptibility ($\chi_o$ = 5.25×10$^{-4}$ emu Oe$^{-1}$ mol- f.u.$^{-1}$) for $Ba_4NbRu_3O_{12}$. The applied field was 1 kOe. The red solid line is the susceptibility fit to the Curie-Weiss law for T from 100 K-300 K. Inset - The magnetization of $Ba_4NbRu_3O_{12}$ as a function of applied field up to 9 T at 1.8 K and 300 K. At 300 K there is a linear relation between the magnetization and the applied field. At low temperature, curvature develops at fields above about 1 T, and there is no sign of magnetic saturation.

Figure 5: Bifurcation of the field cooled (FC) and zero field cooled (ZFC) DC susceptibility in an applied field of 100 Oe for $Ba_4NbRu_3O_{12}$.

Figure 6: Specific heat per mole formula unit of $Ba_4NbRu_3O_{12}$. The full data set is shown in (a) with a fit to the phonon specific heat ($C_{phonon}$) as described in the text. The solid red-line model was constrained to generate a total of $R\ln 2$ magnetic entropy. (b) The change in magnetic entropy $\Delta S_{mag} = \int \left(\frac{C_{mag}(T)}{T}\right) dT$ as inferred by subtracting the phonon model from the measured specific heat: $C_{mag}(T)=C(T)-C_{phonon}(T)$. (c) $C_{mag}(T)$ in the 0.1-100 K regime. The higher $T$ peak indicates buildup of short range magnetic correlations while the lower $T$ peak is associated with spin freezing and the bifurcation point in the FC/ZFC susceptibility. (d) The limiting low T form of $C_{mag}(T)/T$. The small upturn at the very lowest temperatures is associated with the spin-full nuclear isotopes $^{99}$Ru and $^{101}$Ru. Moreover, there exits a finite value of $C_{mag}(T)/T$ as $T \rightarrow 0$.

Figure 7: $C_{mag}(T)$ in a double logarithmic plot. The weak curvature indicates a cross over $T \approx$ 0.4 K between different power-law regimes. Overall, $C_{mag}(T) = \frac{B}{T^2} + \gamma T + AT^\alpha$ accounts well for all data for $T < 3$ K as shown with a solid line. The first term reflects the contribution of the nuclear magnetic moments, the second term is the T-linear contribution to the specific heat, and the third term characterizes the magnetic quasiparticles. These components are shown as dashed lines.

Figure 8: The resistivity of a sintered pellet of $Ba_4NbRu_3O_{12}$ as a function of temperature (inset) and inverse temperature (Main Panel). The data in the elevated temperature regime (250-300 K) was fit to the model $\rho = \rho_o e^{\frac{E_a}{k_b T}}$ (red line) with $E_a = 0.22$ eV.

Figure 9: The HOMOs and LUMOs (Highest energy occupied and lowest energy unoccupied) orbitals for $Ba_4NbRu_3O_{12}$ calculated using extended Hückel theory with relativistic effects included. (There are two types of orbitals at the Fermi energy.) The results for two of the trimers are shown. The three trimers per unit cell are all equivalent by symmetry, but the orbital diagrams are shown for only two of the trimers in this figure for clarity; all the trimers are the same. The unit cell of $Ba_4NbRu_3O_{12}$ is shown for comparison; the dashed outline indicates the primitive unit cell for the MO calculations. The signs of the wave function in the lobes are differentiated by the red and blue colors – the distribution of lobes is typical of an antibonding orbital electron distribution.

Figure 10: (a) The calculated Ab initio electronic band structures (spin orbit coupling included) and (b) The calculated total density of states (DOS) of $Ba_4NbRu_3O_{12}$.

**Table 1. Structural parameters for $Ba_4NbRu_3O_{12}$ at 300 K. Space group $R$-$3m$ (No. 166).**

| Atom | Wyckoff. | Occ. | x | y | z | $U_{iso}$ |
|---|---|---|---|---|---|---|
| Ba1 | 6c | 1 | 0 | 0 | 0.12933(5) | 0.0153(5) |
| Ba2 | 6c | 1 | 0 | 0 | 0.28678(5) | 0.0113(5) |
| Nb | 3a | 1 | 0 | 0 | 0 | 0.0044(9) |
| Ru1 | 3b | 1 | 0 | 0 | ½ | 0.0156(8) |
| Ru2 | 6c | 1 | 0 | 0 | 0.41222(7) | 0.0107(6) |
| O1 | 18h | 1 | 0.4923(6) | 0.5077(6) | 0.1224(2) | 0.022(3) |
| O2 | 18h | 1 | 0.5055(6) | 0.4945(6) | 0.2933(3) | 0.013(2) |

a = 5.75733(6) Å, c = 28.5870(4) Å

$R_{wp}$ = 11.97%, $R_p$ = 8.33%, $R_F^2$ = 5.1%

**Table 2. Parameters determined by fitting two Debye Models and an Einstein Mode to high $T$ specific heat data. $\Theta_{D1}$, $\Theta_{D2}$, and $T_E$ are the corresponding temperature scales.[21] The high $T$ limit of each term is denoted $C_{D1,max}$, $C_{D2,max}$ and $C_{E,max}$. The fit sums to 58(3) $R$, which is consistent with the Dulong-Petit limit of 60 $R$ per formula unit. The error bars are such that the chi-squared parameter of the fit remains within 10% of its original value with variation in each parameter. [21]**

| Parameter | Constrained Fit |
|---|---|
| $C_{D1,max}$ ($R$) | 25.9(5) |
| $\Theta_{D1}$ (K) | 308(2) |
| $C_{D2,max}$ ($R$) | 28(3) |
| $\Theta_{D2}$ (K) | 813(70) |
| $C_{E,max}$ ($R$) | 3.6(1) |
| $T_E$ (K) | 87.2(4) |

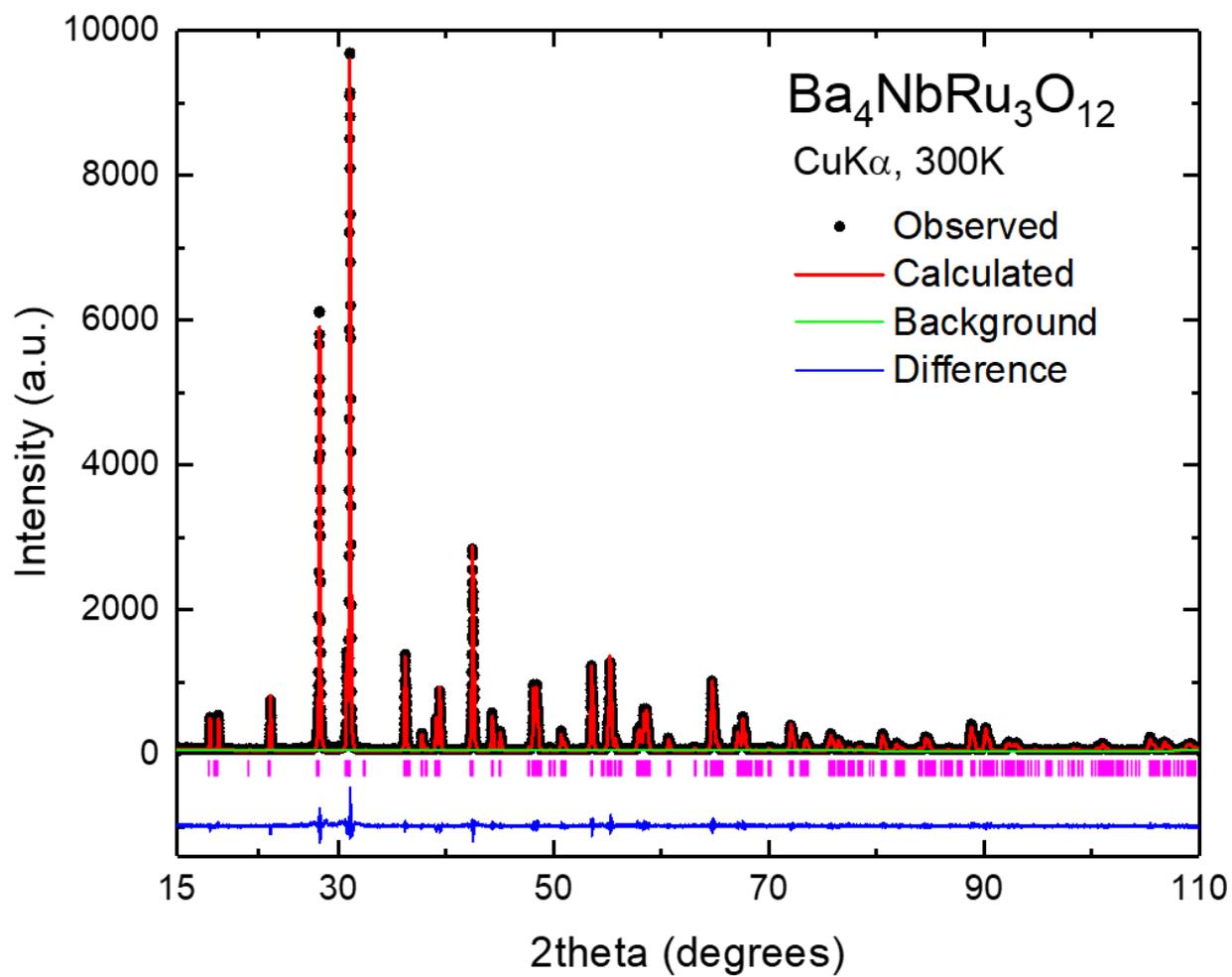

**Figure 1**

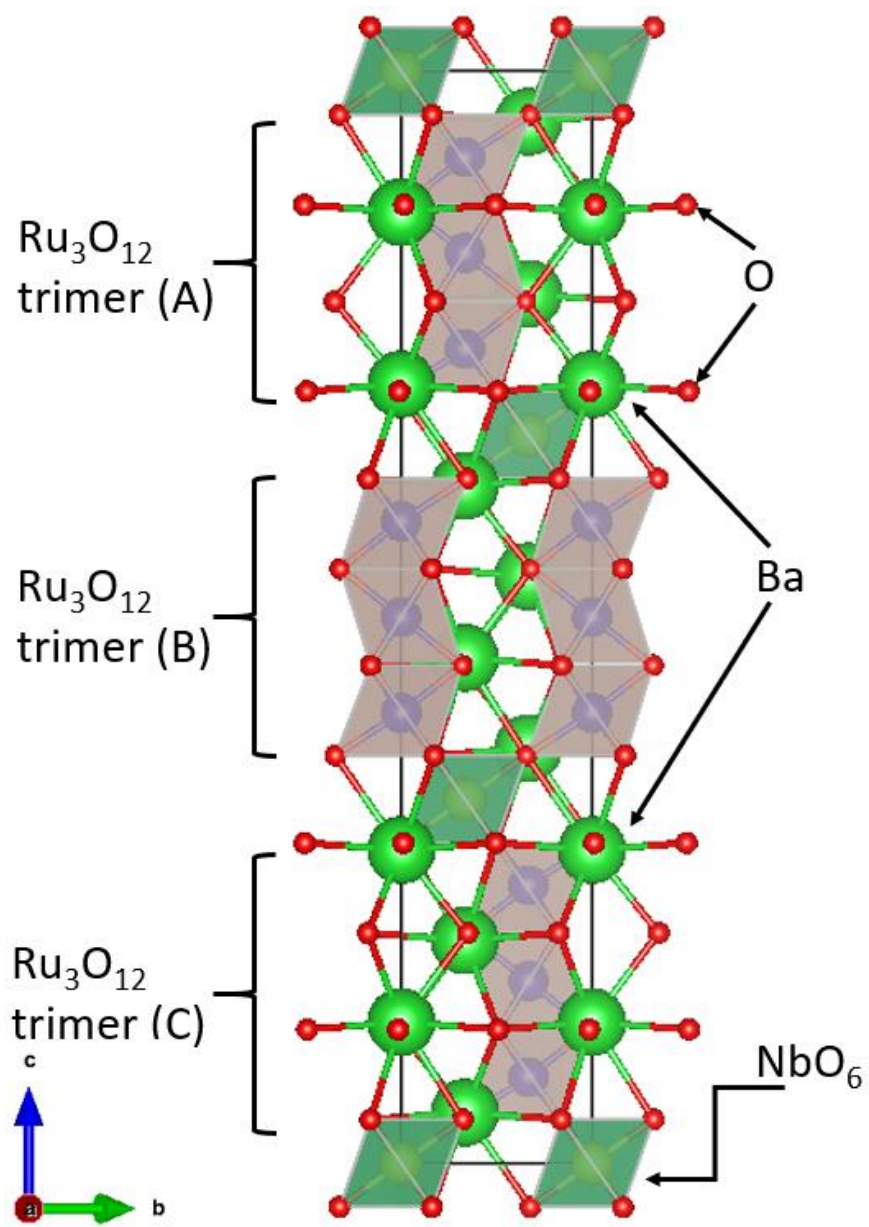

**Figure 2**

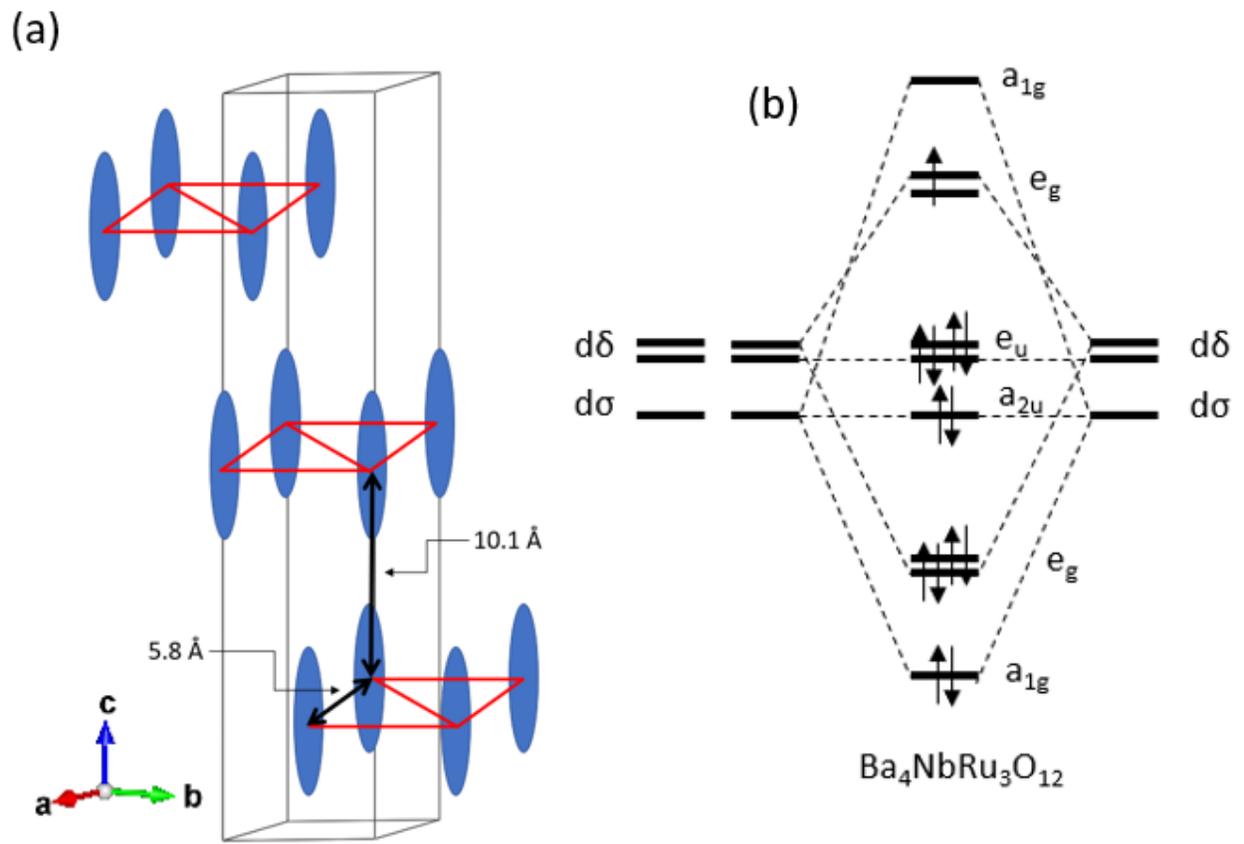

**Figure 3**

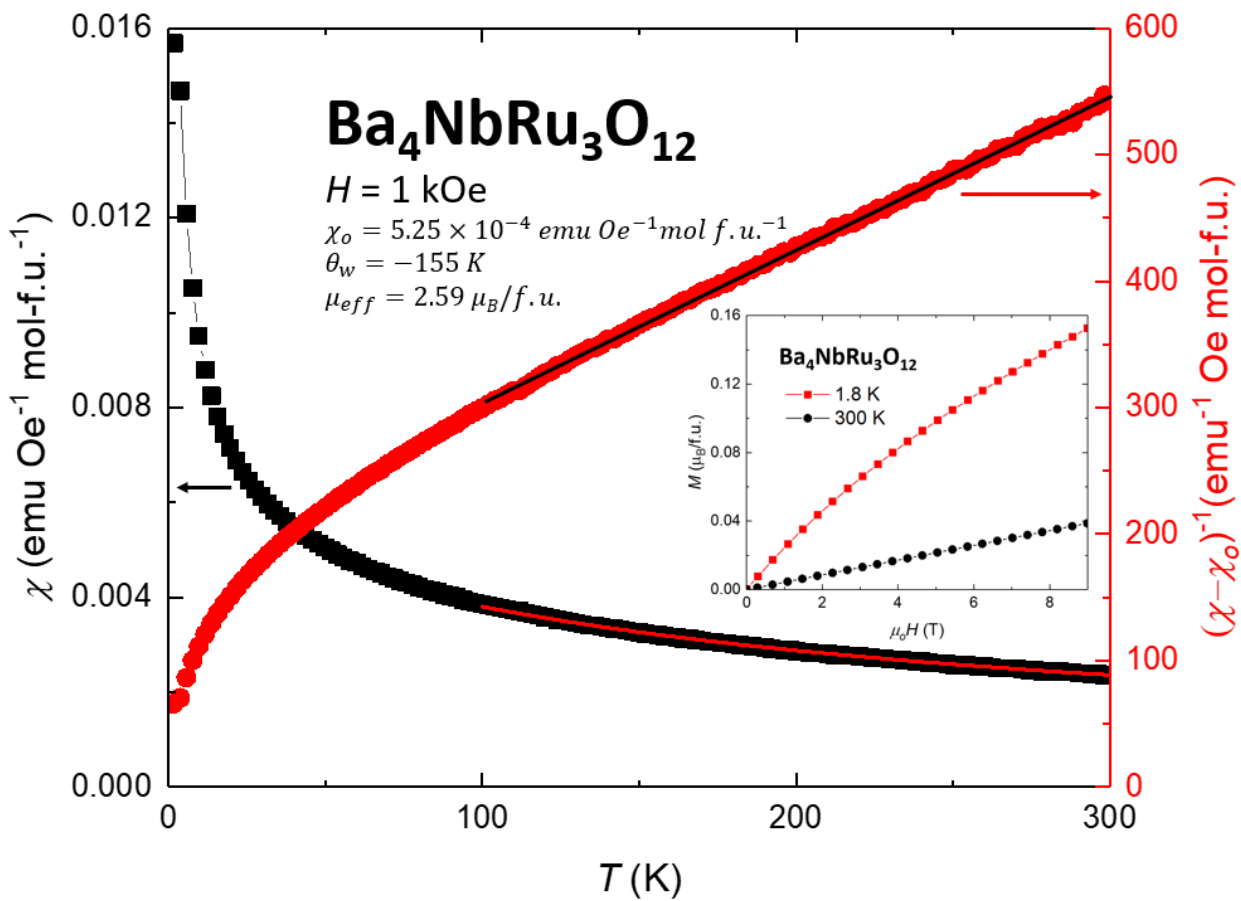

**Figure 4**

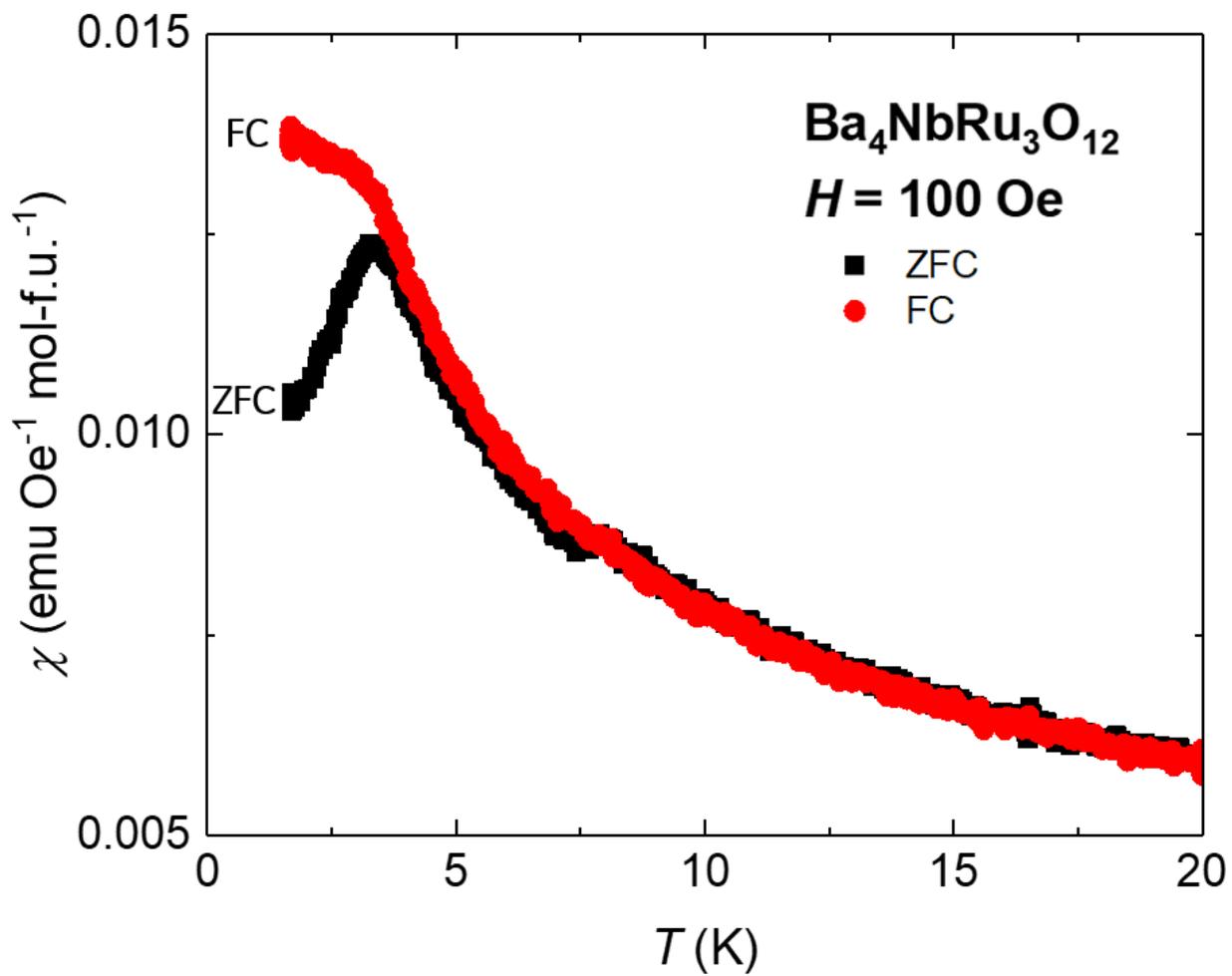

**Figure 5**

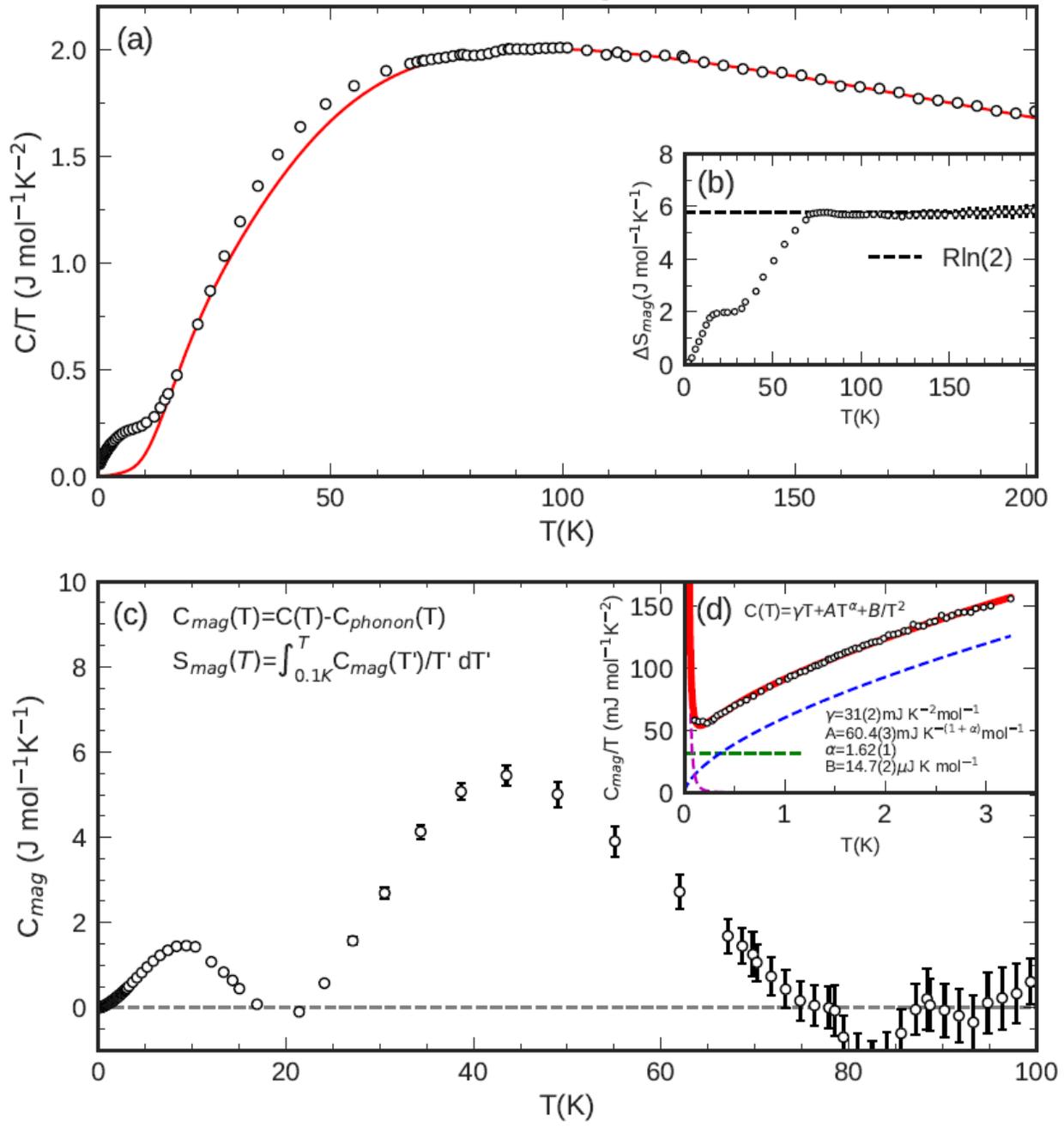

**Figure 6**

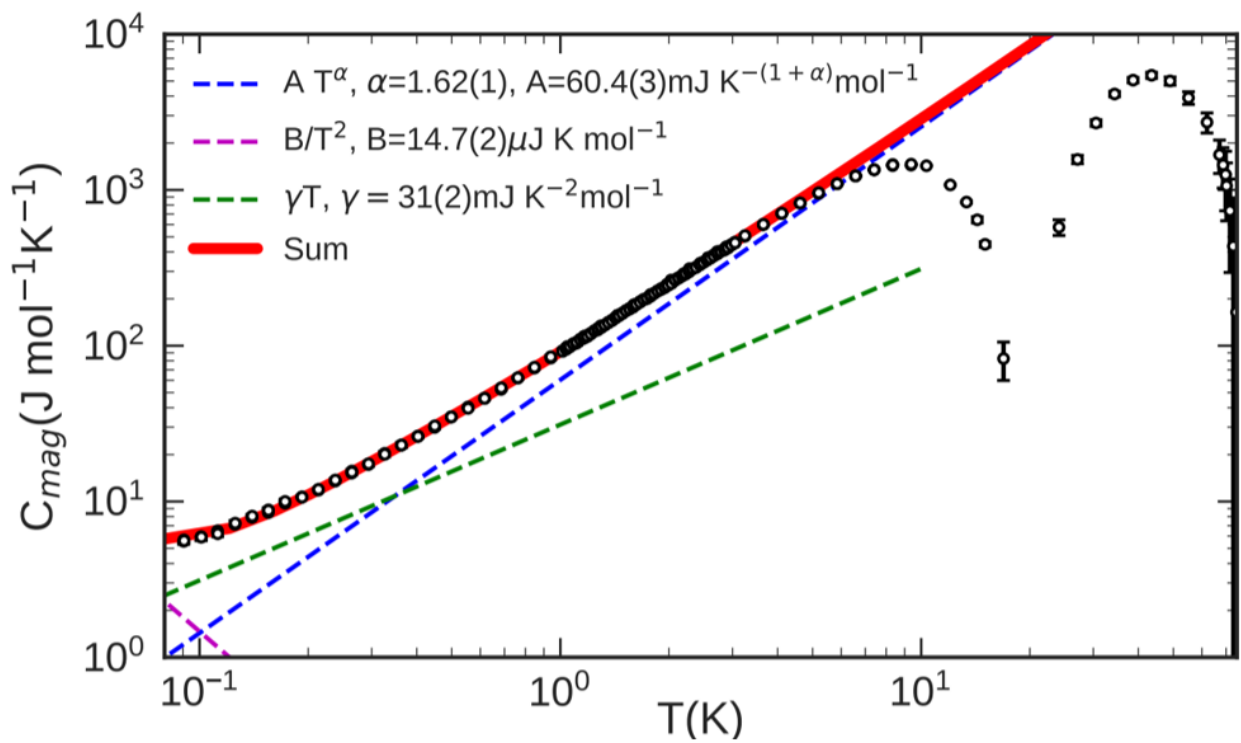

**Figure 7**

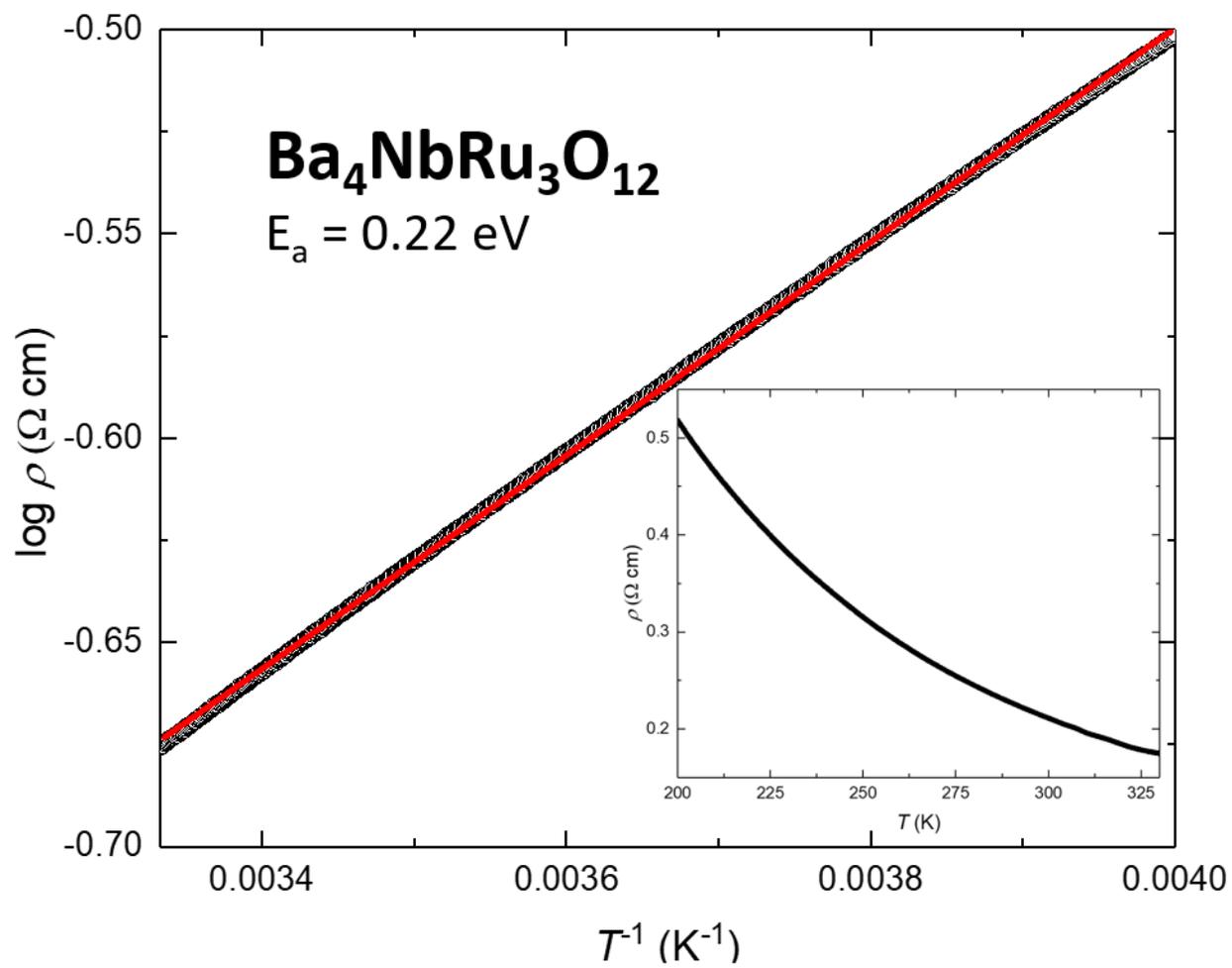

**Figure 8**

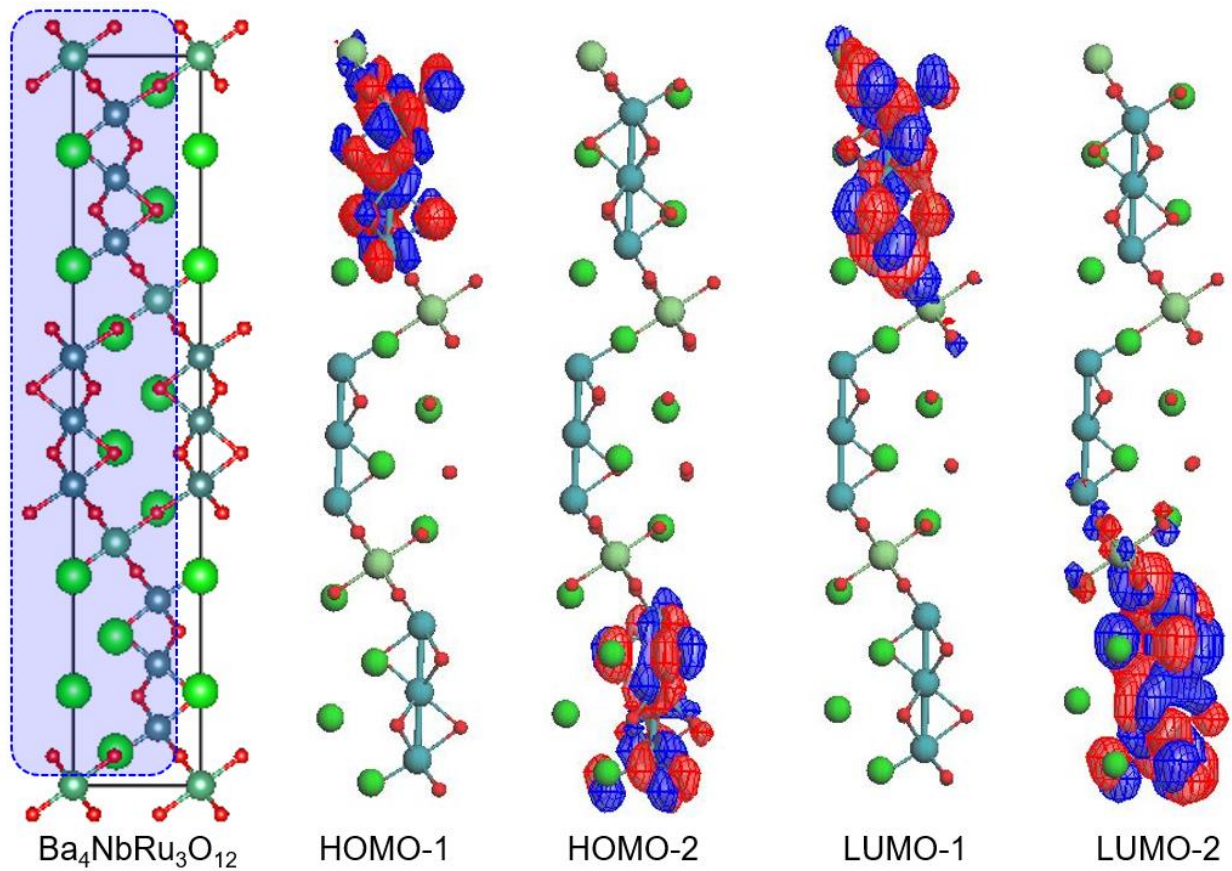

Ba$_4$NbRu$_3$O$_{12}$    HOMO-1    HOMO-2    LUMO-1    LUMO-2

**Figure 9**

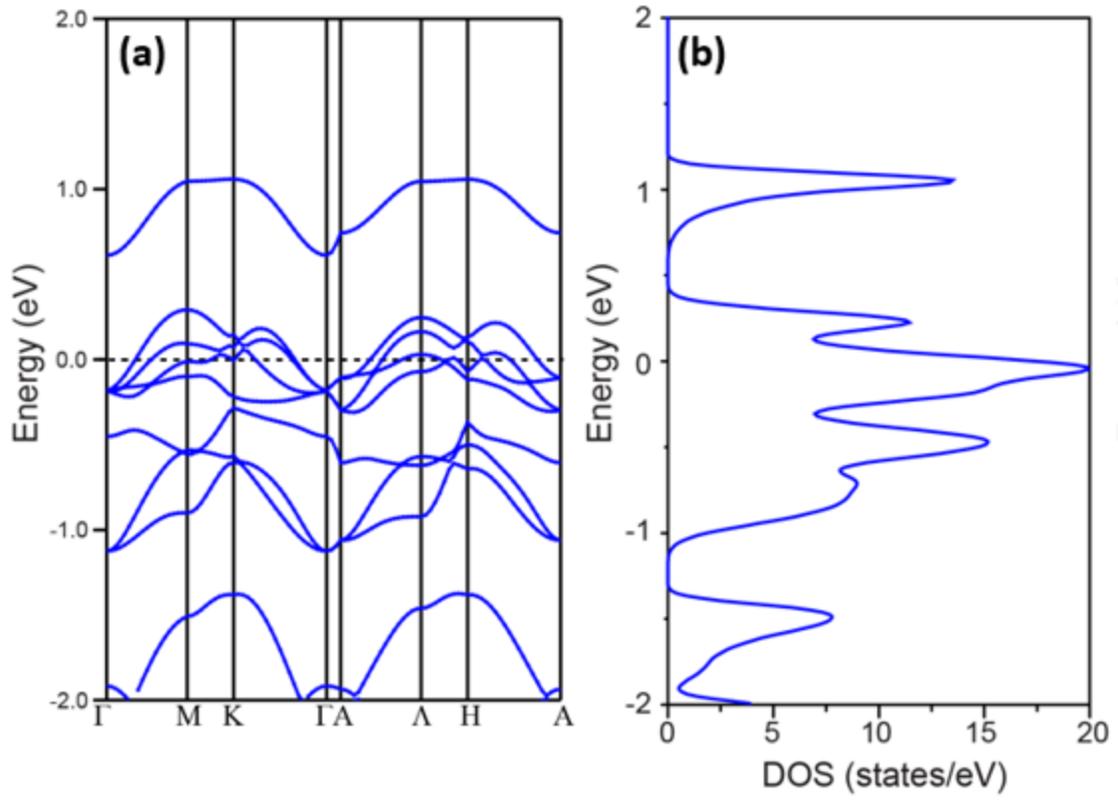

**Figure 10**